# Ultrahigh carrier mobility, Dirac cone and high stretchability in pyrenyl and pyrazinoquinoxaline graphdiyne/graphyne nanosheets confirmed by first-principles

Fazel Shojaei*,a and Bohayra Mortazavi**,b

*a*Department of Chemistry, Faculty of Sciences, Persian Gulf University, Bushehr 75169, Iran.
*b*Chair of Computational Science and Simulation Technology, Institute of Photonics, Department of Mathematics and Physics, Leibniz Universität Hannover, Appelstraße 11,30167 Hannover, Germany.

## Abstract

Graphdiyne nanomaterials are low density and highly porous carbon-based two-dimensional (2D) materials, with outstanding application prospects for electronic and energy storage/conversion systems. In two latest scientific advances, large-area pyrenyl graphdiyne (Pyr-GDY) and pyrazinoquinoxaline graphdiyne (PQ-GDY) nanosheets have been successfully fabricated. As the first theoretical study, herein we conduct first-principles simulations to explore the stability and electronic, optical and mechanical properties of Pyr-GDY, N-Pyr-GDY, PQ-GDY and N-Pyr-GYN monolayers. We particularly examine the intrinsic properties of PQ-graphyne (PQ-GYN) and Pyr-graphyne (Pyr-GYN) monolayers. Acquired results confirm desirable dynamical and thermal stability and high mechanical strength of these novel nanosheets, owing to their strong covalent networks. We show that Pyr-based lattices can show high stretchability. Analysis of optical results also confirm the suitability of Pyr- and PQ-GDY/GYN nanosheets to adsorb in the near-IR, visible, and UV range of light. Notably, PQ-GDY is found to exhibit distorted Dirac cone and highly anisotropic fermi velocities. First-principles results reveal ultrahigh carrier mobilities along the considered nanoporous nanomembranes, particularly PQ-GYN monolayer is predicted to outperform phosphorene and $MoS_2$. Acquired results introduce pyrenyl and pyrazinoquinoxaline graphyne/graphyne as promising candidates to design novel nanoelectronics and energy storage/conversion systems.

## 1. Introduction

Since 2004, graphene [1–3] has been on the focus of extensive researches, owing to its exceptional physics and chemistry. Graphene, the planar form of full-$sp^2$ carbon atoms has been known for more than half of a century before its first synthesis report. The nature of carbon atoms allows them to form different bonding and evolve to diverse structures. Such that despite the superior stability of graphene, producible atomic-layer thick form of carbon atoms was originally known not to be limited to this full-$sp^2$ lattice. In fact, more than three decades ago, Baughman *et. al* [4] predicted numerous forms of hybrid sp and $sp^2$ carbon nanoporous 2D lattices, so called graphdiyne/graphyne. These nanoporous structures mostly include benzene rings connected by acetylene bonds. Depending on the architecture of core atoms and number of carbon atoms in the connecting carbon chains, different lattices can be formed. In this regard, graphyne and graphdiyne are distinguishable by the number of the carbon atoms in the connecting acetylene chains, including either two or four carbon atoms, respectively.

Despite distinctly lower stability, mechanical strength and thermal conductivity and complexity of fabrication of graphdiyne/graphyne lattices in comparison with graphene, they were known to offer highly promising properties that graphene was unable to deliver in its pristine form. Among them, these carbon-based nanoporous structures are mostly semiconductors [5–12], whereas graphene does not show an electronic band gap. Presenting a suitable band gap is a critical requirement for the majority of advanced applications in nanoelectronics, optoelectronics and catalysis. Moreover, defect free and densely packed structure of graphene results in brittle deformation due to the limited stretchability of carbon-carbon bonds. In contrast graphdiyne/graphyne porous lattices allow the remarkable flexibility and in some cases they can show polymer-like superstretchability [7,13]. Highly porous structure also facilitate the access and reaction in the active sites of graphdiyne/graphyne lattices, which has been proven to boost the efficiency for the applications in rechargeable batteries [14–17], catalysts [18] and hydrogen storage [19,20] systems. Notably, three orders of magnitude lower thermal conductivity of graphdiyne lattices [13,21] in comparison with graphene, offer them as unique candidates to design light carbon-based thermoelectric devices [22,23] and thermal insulators.

Nonetheless, unlike graphene which is naturally available in the multi-layer form of graphite, graphdiyne/graphyne structures require a more elaborated fabrication process. The unique properties and broad application prospects of graphdiyne/graphyne nanomembranes however have been serving as strong motives for designing novel synthesis routes. While to date there has been no report for synthesis of graphyne nanosheets, during the past decade exciting experimental advances have been accomplishment with respect to the fabrication of graphdiyne nanosheets. In 2010, Li *et al.* [24] reported the first experimental realization of graphdiyne by using a cross-coupling reaction utilizing the hexaethynylbenzenethe monomer. It was in 2017 that Matsuoka *et al.* [25] reported the experimental realization of graphdiyne nanomembranes at gas/liquid or liquid/liquid interface. This experimental advance paved the path for the synthesis of novel large-scale graphdiyne structures. Shortly after, Kan *et al.* [26] and Wang *et al.* [27] succeeded in the fabrication of crystalline nitrogen-graphdiyne and boron-graphdiyne nanosheets, respectively. Using the concept devised originally by Matsuoka *et al.* [25] in 2017, the same group later reported the experimental realization of triphenylene graphdiyne via employing hexaethynyltriphenylene monomer [28]. In line with continuous experimental advances in the design and synthesis of novel graphdiyne structures, Wang *et al.* [29] most recently reported the first realization of large-area pyrenyl graphdiyne (Pyr-GDY) monolayer via van der Waals epitaxial growth over hexagonal boron nitride. Shortly after, Gao and coworkers [30] reported the successful synthesis of high quality pyrazinoquinoxaline graphdiyne using a bottom-up chemical synthesis. Motivated by these latest exciting accomplishments, in this study our objective is to examine the stability, optical and electronic features and mechanical response of pyrenyl and pyrazinoquinoxaline graphdiyne/graphyne using the extensive density functional theory (DFT) calculations. In particular, we also consider the lattices made of nitrogen doped pyrenyl groups.

## 2. Computational methods

In this work, all structure optimizations, thermal stability assessments, and electronic structure calculations are performed using density functional theory (DFT) within the framework of generalized gradient approximation (GGA) and Perdew–Burke–Ernzerhof (PBE) [31], as implemented in *Vienna Ab-initio Simulation Package* [32,33]. Projector augmented wave method was used to treat the electron-ion interactions [34,35]. For geometry optimizations, atoms and lattices were relaxed according to the Hellman-Feynman forces using conjugate gradient algorithm with and energy cut off of 500 eV until forces were lower than 0.002 eV/Å [36]. The first Brillouin zone (BZ) was sampled with Γ-centered k-gird of 8×8×1 for rhombic lattices and 8×4×1 for rectangular lattices. The complete mechanical properties are examined by conducting uniaxial tensile simulations. Since GGA-PBE systematically underestimates the band gaps, HSE06 hybrid functional was employed for high-accuracy electronic and optical properties calculations [37]. Ab-initio molecular dynamics (AIMD) simulations were conducted over rectangular unitcells with a time step of 1 ps. Moment tensor potentials (MTPs)[38] are trained to evaluate the interatomic forces [39]. In our previous work [39] it was confirmed that the MTPs can be employed to accurately examine the phonon dispersion relations, dynamical stability, group velocity and various thermal properties. The training sets for the development of MTPs are prepared by conducting AIMD simulations for 1500 time steps, in which the temperature was gradually increased from 20 to 100 K. The phonon dispersions were then obtained for rectangular unitcell using the PHONOPY code [40] with fitted MTPs for interatomic force calculations [39] over 5×5×1 supercells.

The mobility of a charge carrier (electron or hole) in a pristine defect-free semiconductor/insulator is dominated by the interaction of the charge carrier with acoustic phonons of the lattice. We used the deformation potential theory (DPT), originally developed by Bardeen and Shockley [41], to calculate the acoustic phonon-limited charge carrier mobility within considered nanosheets. The theory is originally formulated for isotropic materials, however because of the structural anisotropy of our nanosheets, we used an adjusted version of DPT to take account for the anisotropy. In anisotropy-corrected DPT, electron and hole mobilities along in-plane directions (zigzag ($\mu_z$) and armchair ($\mu_a$)) are calculated via [42–44]:

$$\mu_z = \frac{5e\hbar^3(5C_z+3C_a)}{2k_bT m_z^{*\frac{3}{2}} m_a^{*\frac{1}{2}}(9E_z^2+7E_zE_a+4E_a^2)} \quad (1)$$

$$\mu_a = \frac{5e\hbar^3(5C_a+3C_z)}{2k_bT m_a^{*\frac{3}{2}} m_z^{*\frac{1}{2}}(9E_a^2+7E_aE_z+4E_z^2)} \quad (2)$$

in which $\hbar$ is the reduced Planck constant, $k_b$ is the Boltzmann constant, $C$ and $m^*$ are the elastic modulus and the effective mass, respectively, and $E$ is the deformation energy constant and it mimics the charge carrier-phonons interactions for the i[th]-states along the transport direction. The charge carrier mobility can be further suppressed through scattering by optical phonons, defects, impurities, and crystal zone boundaries. Therefore, the calculated mobilities using (1) and (2) are upper limits to the experimental values.

We examined the light absorption properties of each system by calculating their frequency-dependent dielectric matrix, neglecting the local field effects. The imaginary part ($\varepsilon_2$) of the frequency-dependent dielectric matrix is obtained from the following equation:

$$\varepsilon_{\alpha\beta}^{2}(\omega) = \frac{4\pi^2 e^2}{\Omega} \lim_{q \to 0} \frac{1}{q^2} \sum_{c,v,\mathbf{k}} 2w_k \delta(\varepsilon_{c\mathbf{k}} - \varepsilon_{v\mathbf{k}} - \omega) <u_{c\mathbf{k}+\mathbf{e}_\alpha q} | u_{v\mathbf{k}}><u_{c\mathbf{k}+\mathbf{e}_\beta q} | u_{v\mathbf{k}}>^* \quad (3)$$

where indices $c$ and $v$ refer to conduction and valence band states, respectively; $w_k$ is the weight of the $k$-point; and $u_{ck}$ is the cell periodic part of the orbitals at the $k$-point. The real part ($\varepsilon_1$) of the tensor is obtained from the Kramers-Kronig relation [45]. The absorption coefficient is calculated from the following:

$$\alpha(\omega) = \sqrt{2}\omega \left[\frac{\sqrt{\varepsilon_1^2 + \varepsilon_2^2} - \varepsilon_1}{2}\right]^{1/2} \quad (4)$$

## 3. Results and discussion

In this section we first describe the structural features of these novel graphyne (GYN) and graphdiyne (GDY) structures. Fig. 1 shows the molecular geometry of the three π-conjugated core molecules of pyrene (Pyr), pyridinic N-doped pyrene (N-Pyr), and pyrazinoquinoxaline (PQ) as well as the crystal structure of these molecules-based GYN and GDY nanosheets. The main structural properties of each sheet is also summarized in Fig. 1. The figure also shows that all these 2D networks possess a centered rectangular lattice symmetry and a rhombic primitive cell with Pm (No. 6) space group. It can be seen that similar to the π-GYN/GDY, PQ-GYN/GDY also contains six acetylenic linkages, while Pyr-GYN/GDY and N-Pyr-GYN/GDY contain only four ones. The length of ≡C-C≡ single-like bonds and -C≡C- triple-like bonds in acetylenic linkages are found to be 1.34 and 1.22-1.23 Å, respectively, which are almost identical to the previously reported values for graphdiyne [6,46]. The geometric coordinates of stress free structures in the VASP POSCAR format are included in given in the supplementary information to facilitate the upcoming studies.

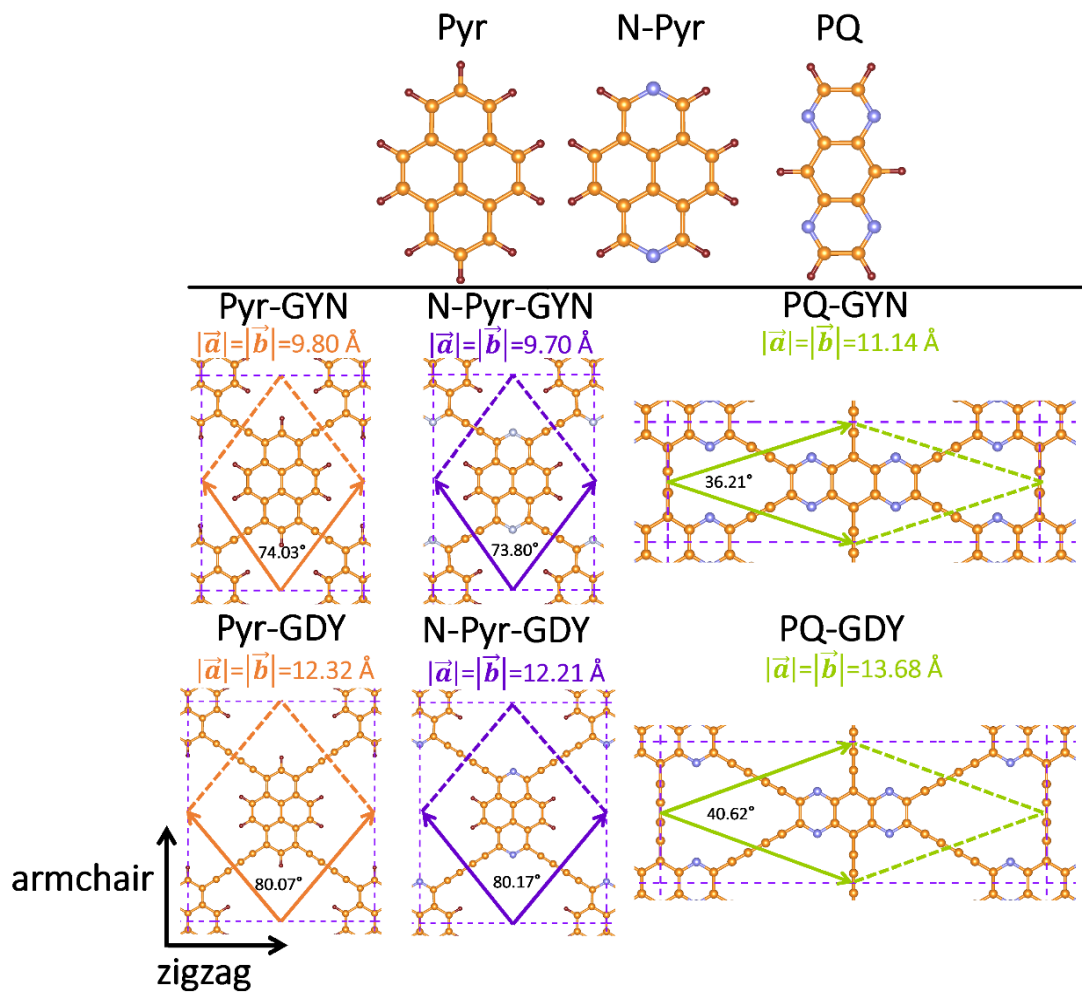

**Fig. 1**, Schematic of pyrene (Pyr), pyridinic N-doped pyrene (N-Pyr), and pyrazinoquinoxaline (PQ) molecules as well as the crystal structure for these molecules-based GYN and GDY in rhombic and rectangular lattices. The structural parameters are also shown for each nanosheet. Orange, blue, and dark brown circles represent carbon, nitrogen, and hydrogen atoms.

We first examine dynamical stability by calculating the phonon dispersion relations. The acquired phonon dispersion relations for the considered monolayers up to 30 and 100 THz frequencies are illustrated in Fig. 2 and Fig. S1, respectively. Like other 2D materials, all these monolayers show three acoustic modes starting from the Γ point, two with linear dispersions and the other one with quadratic relation. As it is conspicuous, the phonon dispersion relations are free of imaginary frequencies and thus confirming desirable dynamical stability of considered monolayers. To assess the thermal stability of Pyr-, N-Pyr-, and PQ-GYN/GDY, we performed ab initio molecular dynamics (AIMD) using rectangular supercells at two temperatures of 1000 and 3000 K for 20 ps with a time step of 1 fs. As shown in Fig. S2, at 1000 K, the crystalline integrity and general characteristics of each monolayer are retained after 20 ps and their potential energies fluctuate around certain average magnitudes, approving their desirable thermal stabilities. The figure also indicates that Pyr- and N-Pyr-based nanosheets exhibit excellent stability at ultrahigh temperature of 3000 K, while the PQ-based systems disintegrate at this temperature. The potential energy profiles of PQ-GYN and PQ-GDY have a step at 19 and 14 ps, corresponding to multiple covalent bonds breaking processes. These results confirm remarkable thermal study of these novel 2D systems, promising for the high temperature applications.

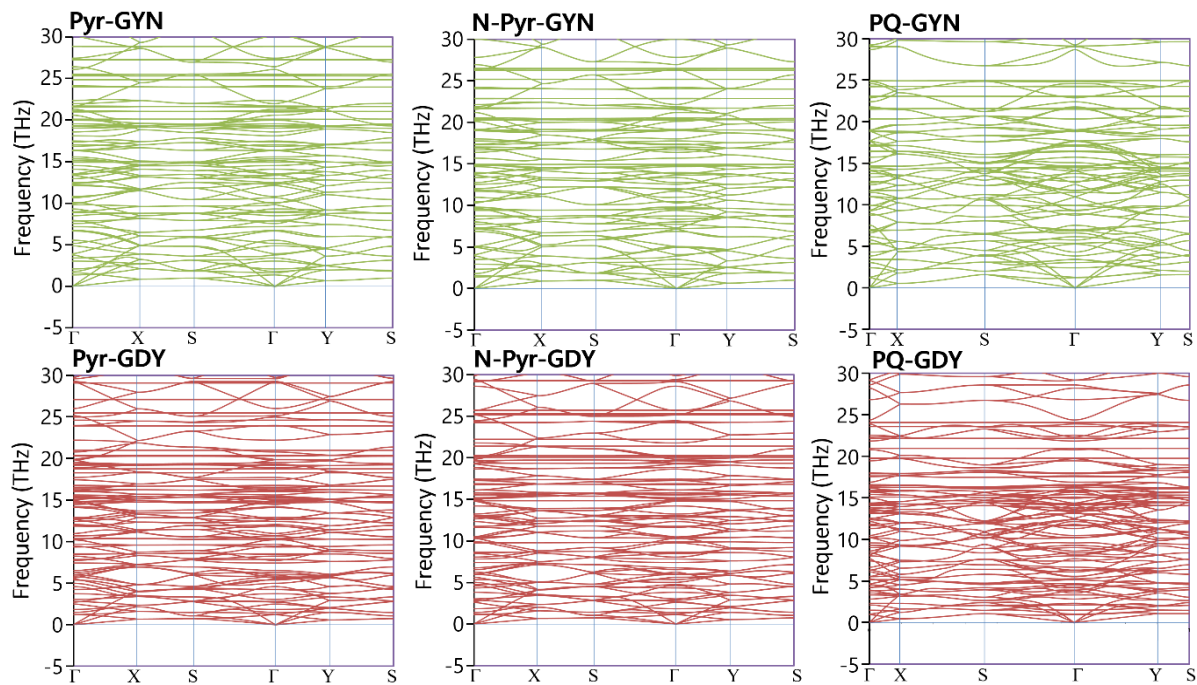

**Fig. 2**, Phonon dispersion relations acquired using the MTP with PHONOPY code over 5×5×1 rectangular supercells [39]. The complete spectrum is presented in Fig. S1.

We next evaluate the mechanical properties of these studied carbon-based nanomaterials using the uniaxial stress-strain relations. In these calculations the stresses along the two perpendicular directions of the loading are ensured to stay negligible. Along the normal direction of the sheets this goal is achieved automatically upon the geometry optimization due to the contact with vacuum. For the other planar direction, the stress is reached to a negligible value by altering the periodic box size [47]. Herein the mechanical responses are evaluated along the armchair and zigzag direction to examine the anisotropy. The predicted uniaxial stress-strain relations are depicted in Fig. 3. As it is clear, for the all considered structures along the armchair loading direction the monolayers can extend to higher strain levels. Unlike the conventional materials, for which the stress-strain relation shows an initial linear part, our results reveal that Pyr-based lattices exhibit no clear linear elasticity. For the aforementioned nanosheets, the stress-strain relations start with fluctuating patterns, followed by a linear pattern and leading to a non-linear pattern before reaching the maximal tensile strength point. PQ-based nanomembranes however show the conventional behavior, which includes linear elasticity followed by a non-linear relation. For the Pyr-based lattices, the tensile strengths for GDY and GYN lattices are very close, around 18 N/m and 12 N/m along the armchair and zigzag directions, respectively. Nonetheless, Pyr-GYN nanosheets show generally lower stretchabilities than GDY counterparts. These results suggest that doping the original Pyr group with N-Pyr does not lead to substantial changes in the mechanical response. On the other hand, for the case of PQ-based nanomembranes, GYN lattice shows distinctly higher elastic modulus and tensile strengths than GDY counterparts, in accordance with the classical theory expecting higher mechanical properties by increasing density. The maximum tensile strength of PQ-GYN and PQ-GDY are predicted to be 24.2 and 11.9 N/m, respectively, both occurring along the zigzag direction. Notably, the tensile strength of PQ-GYN is only by around 40% lower than that of the ultrahigh strong graphene [48]. The elastic modulus of PQ-GYN along the armchair and zigzag are predicted to be 100 and 228 N/m, respectively, distinctly higher than the corresponding values of 89 and 169 N/m, respectively for the PQ-GDY.

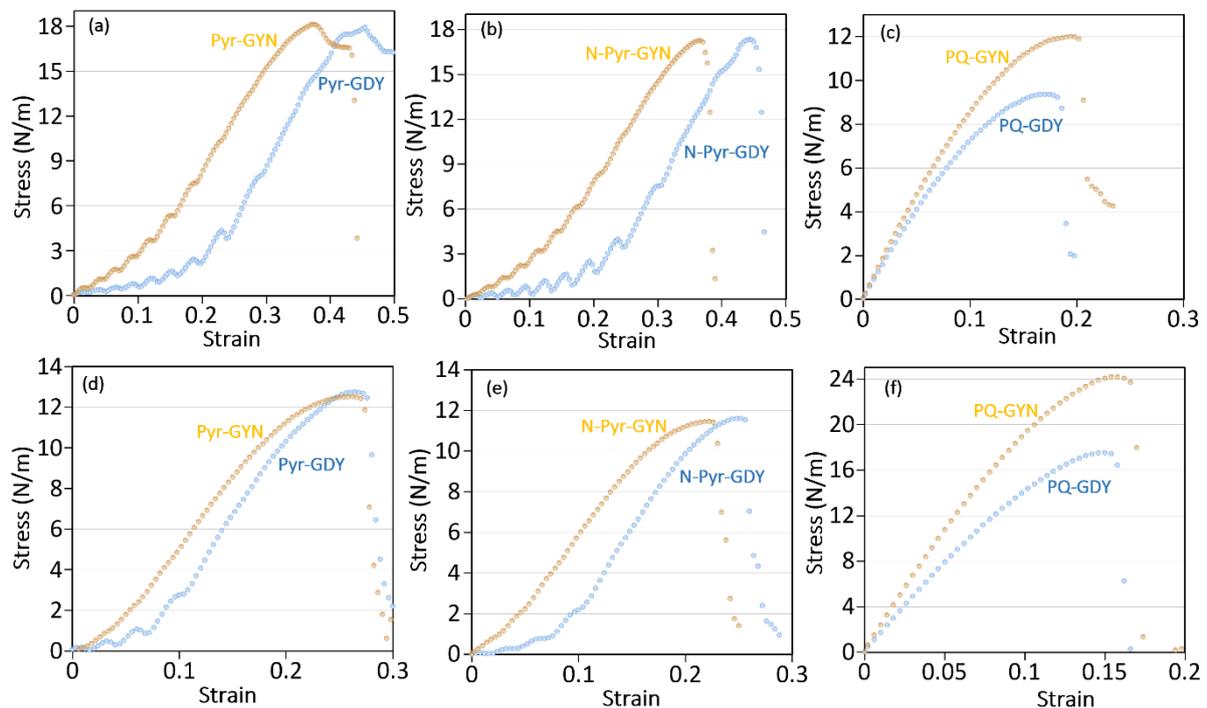

**Fig. 3**, Uniaxial stress-strain responses along the (a, b and c) armchair and zigzag (d, e and f) directions.

In order to understand the unusual mechanical properties of considered nanoporous structures, the mechanism of load transfer in should be taken into consideration. For the conventional materials with densely packed lattice, like diamond and graphene, the elongation during the loading can be only achieved by direct increasing of the bond lengths. Due to the harmonic nature of bond elongations, at low strain levels the stress increases linearly, resulting in a linear relation. For graphdiyne structures, because of the nanoporousity of the lattices, at low strain levels the deformation can be occurred via a combination of bond elongation and structural deflection [7]. Let's consider Pyr-GDY structures at different strain levels, as illustrated in Fig. 4. As already shown in Fig. 3a and 3b, Pyr-GDY and N-Pyr-GDY monolayers show remarkable fluctuations in the uniaxial stress-strain relations for the loading along the armchair direction. It is clear that at remarkably high strain levels, Pyr-based lattices contract substantially along the width (find Fig. 4b and 4e). It is conspicuous that the porosity of lattice provides free space for the carbon chains to rotate and further orient along the loading, resulting in a rubber-like behavior. For the same lattices loaded along the zigzag direction, the carbon chains also easily rotate but become very early oriented along the loading (find Fig. 4c and 3f) direction. In these cases, despite the high available porosity, the structure is with high degree oriented along the loading and such that the further stretching of the structure is mostly accomplished by the bon elongation, rather than the sheet contraction along the width. This way, the lattice structure and the possibility of carbon chains for rotation and contraction of the sheet along the width can result in enhanced superstretchability. Since the elongation of bonds are limited, the enhancement of stretchability can be achieved by improving the possibility of structural deflections, which results in vanishing the conventional linear elasticity. To better understand this fact, one should consider PQ-GDY lattice, in which the existence of four carbon chains in a rectangular unitcell limit their rotation and results in considerably lower stretchability in comparison with Pyr-based counterparts. Our results also reveal similar failure mechanism for Pyr-GDY and N-Pyr-GDY nanosheets, which confirm that N-doping does not affect the mechanical behavior. For the aforementioned lattices along the armchair direction, the rupture occurs in the central benzene rings (find Fig. 4b and 4e), whereas along the zigzag direction the rupture occurs in the bond connecting the central benzene rings to the carbon chains (find Fig. 4f and 4d).

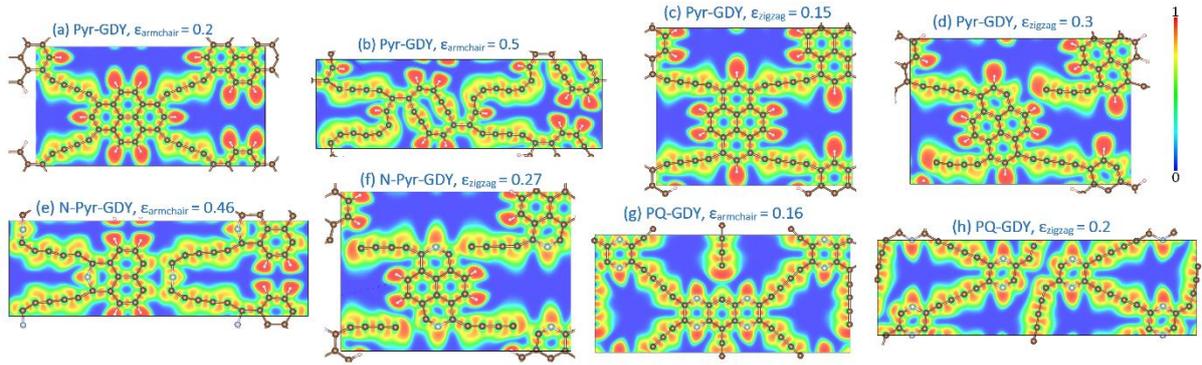

**Fig. 4**, Top views of the deformation process of graphdiyne nanosheets at strain levels (ε) along the armchair and zigzag direction. Contour illustrates the electron localization function (ELF) within the unit-cell.

We now shift our attention to the evaluation of electronic properties of these novel 2D carbon-based networks. Fig. 5 shows the HSE06 band structures of Pyr-, N-Pyr-, and PQ-GYN/GDY monolayers along their high symmetry paths in the first Brillouin zone (FBZ). For each monolayer the partial charge densities of valance band maximum (VBM) and conduction band minimum (CBM) are also calculated and shown next to its band structure. From the illustration, it is evident that the monolayers' electronic structures sensitively vary with the type of core molecule (Pyr and PQ) and length of acetylenic linkage (GYN and GDY). However, because Pyr and N-Pyr share several structural features, achieving almost similar band structures for pairs of Pyr-GYN and N-Pyr-GYN or Pyr-GDY and N-Pyr-GDY is not unexpected. The four Pyr- and N-Pyr-based monolayers are direct gap semiconductors with HSE06 (PBE/GGA) band gaps of 0.84 (0.44) eV for Pyr-GYN, 0.88 (0.46) eV for N-Pyr-GYN, 0.92 (0.51) eV for Pyr-GDY, and 0.94 (0.52) eV for N-Pyr-GDY. The band gap occurs at C point for Pyr- and N-Pyr-GYN, while it occurs at Γ point for Pyr- and N-Pyr-GDY. The HSE06 band gap of Pyr-GDY is appreciably smaller than the experimental value (2.07 eV) [49]. A similar observation has been also reported for the monolayer γ-GDY, in which its HSE06 band gap (0.89 eV) is even much smaller than the experimental value obtained for multilayer GDY film (1.40 eV) [50]. Using GW method to obtain an accurate band gap did not solve this data discrepancy, because the predicted GW band gap (1.10 eV) for γ-GDY is also smaller than that of the multilayer GDY [51]. The inconsistency between theoretical and experimentally estimated band gaps can be explained due to presence of excessive defects in the crystal structure of stacked layers and also special confinement of limited grain size in the film [49]. Interestingly, although Pyr and N-Pyr have larger π-deloclaization as compared to the hexaring (C6) in γ-GYN/GDY, the magnitude of band gaps obtained for Pyr- and N-Pyr-GYN/GDY are only slightly different than those of γ-GYN (0.94 (0.46) eV) and γ-GDY (0.89 (0.48) eV), calculated at the same level of theory [52]. This can be simply explained through the fact that Pyr- and N-Pyr-GYN/GDY contain lesser number of acetylenic linkages (four) which induces stronger 2D confinement and therefore a noticeable band gap. Our findings, however indicate that PQ-based GYN/GDY with six acetylenic linkages in their primitive cells, exhibit much smaller band gaps than γ-GYN/GDY. According to Fig. 5, PQ-GYN is a direct gap semiconductor with small band gap of 0.17 (0.08) eV at X point, while PQ-GDY is found to be a semimetal with zero band gap by both HSE06 and PBE/GGA functionals. In PQ-GDY's electronic band structure, valence and conduction bands meet at the Fermi level to form a distorted Dirac cone at an off-symmetry point between C and Γ ($\kappa_x/2 + 0s.08\ \kappa_y/2$) in FBZ. The Dirac cone of PQ-GDY in more details is shown in the bottom of Fig. 5. As it can be seen two Dirac points exist in FBZ of PQ-GDY and they are identical due to the centrosymmetry of FBZ. The band dispersions around the Fermi level exhibit a linear but moderately anisotropic behavior, resulting in different Fermi velocities for charge carriers. The fermi velocities are found to be $8.37\times10^4$ m/s along $\kappa_x$ direction and $2.20\times10^5$ m/s (for positive band slope) and $2.92\times10^5$ m/s (for negative band slope) along $\kappa_y$ direction. The Fermi velocities along the $\kappa_y$ are comparable with that of graphene ($8.22\times10^5$ m/s)[53]. It is worth noting that observing Dirac cone band dispersion is very common in graphyne's family. Several other graphynes such as α-, β-, and δ-graphynes have been also theoretically predicted to have Dirac cones in their band structures [54,55].

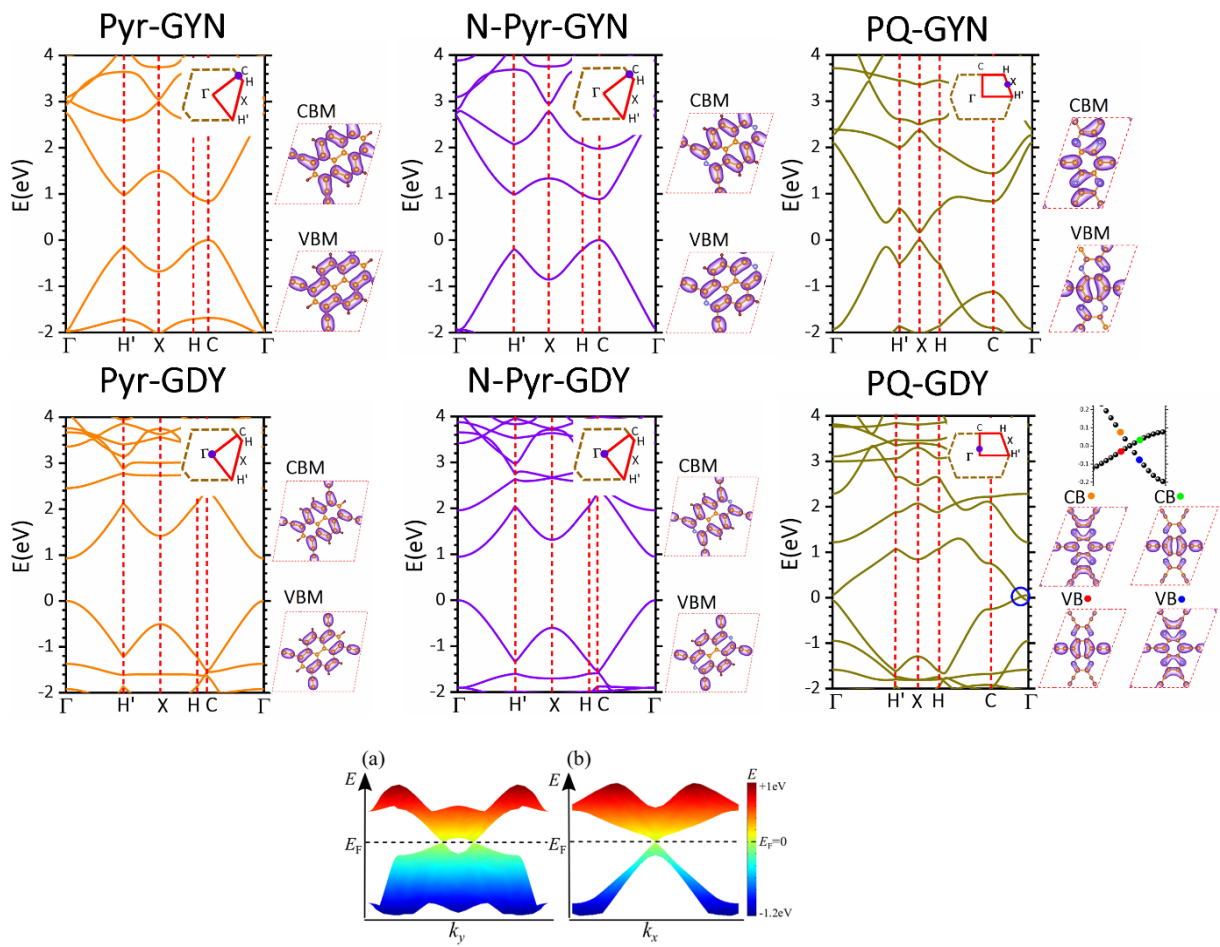

**Fig. 5**, HSE06 electronic band structures of Pyr-, N-Pyr-, and PQ-GYN/GDY monolayers. The partial charge density distributions of VBM and CBM of semiconducting monolayers and that of the frontier states near the Fermi level for the semimetallic one are also shown next to the band structures. For each system the first Brillouin zone (FBZ), points of special symmetry, selected high-symmetry path through the BZ, and the point band gap occurs at are also shown. For PQ-GDY, the 3D band structure of valance and conduction bands are also shown ((a) and (b)). For all systems, the isosurface value for partial charge densities are set to 0.0015 e/Å$^3$.

To explore the origin of conductance valance bands around the Fermi level, we carefully analyze the partial charge densities of VBM and CBM of each monolayer. In all cases, both VBM and CBM are made of almost exclusively out of plane $p_z$ orbitals from the core molecule and acetylenic linkages, in a way that VBM contains π-bonding ($p_z$) states of -C≡C- of the linkages, while CBM contains antibonding π*($p_z$) state of the same -C≡C- bonds. These states are delocalized and distributed over the entire 2D network. The strong interaction between π-conjugation network of core molecules, facilitated through acetylenic linkages, simply explains observing highly dispersed VB and CB in all these monolayers, resulting in very small charge carrier effective masses and high mobilities. For PQ-GDY, it is clear that the occupied states are inversed at the two sides of the Dirac point, which means the energy band inversion.

Next, we investigate the charge carrier mobilities within Pyr-, N-Pyr-GYN/GDY and PQ-GYN along zigzag and armchair directions using effective mass approximation and anisotropy-corrected DPT method. The rectangular cells shown in Fig. 1 are used to calculate the charge carrier mobilities. The corresponding PBE/GGA electronic band structures are also depicted in Fig. S3. Table 1 lists the elastic modulus ($C_{2D}$), effective masses ($m_e^*$, $m_h^*$) of electrons and holes, deformation energy of the CBM and VBM ($E_1^{CBM}$ and $E_1^{VBM}$), and mobility ($\mu_e$, $\mu_h$) of electrons and holes, respectively, along zigzag and armchair directions. Looking at the table, it is apparent that in all cases the magnitudes of effective masses and the deformation energies for both VBM and CBM are different along zigzag and armchair directions. This leads to highly anisotropic electron and hole mobilities within these systems. As an example, for Pyr-GDY, the electron mobility along zigzag direction is found to be 2142.2 $cm^2V^{-1}s^{-1}$, which is by two times smaller than that along the armchair direction (4146.3 $cm^2V^{-1}s^{-1}$). As similar trend is also observed for its hole mobility. The hole mobility along zigzag direction is estimated to be 1385.6 $cm^2V^{-1}s^{-1}$ and this is by six times smaller than that along the armchair direction (8081.9 $cm^2V^{-1}s^{-1}$). It is evident that the hole mobility in Pyr-GDY exhibits higher anisotropy than the electron mobility. These observations qualitatively hold for the other three Pyr-, N-Pyr-based nanosheets. Table 1 also indicates that charge carrier mobility sensitively varies with the core molecule. Interestingly, PQ-GYN is found to exhibit extremely higher electron and hole mobilities of 115708.9 and 59004.7 $cm^2V^{-1}s^{-1}$, respectively, which are notably distinctly larger than the that theoretically estimated for monolayer phosphorene (10000 $cm^2V^{-1}s^{-1}$) and $MoS_2$ (200 $cm^2V^{-1}s^{-1}$) [56,57]. The calculated electron mobility for PQ-GYN is by about two times smaller than those reported for graphene (240000 $cm^2V^{-1}s^{-1}$) [58,59] and γ-GDY (208100 $cm^2V^{-1}s^{-1}$)[60], obtained using the same theory. The comparison of the highest electron and hole mobilites in each system indicates that because of the Pyr- and N-Pyr-nanosheets with higher hole mobilities can exhibit p-type semiconducting properties, while PQ-GYN with dramatically larger electron mobility is expected to act as a n-type semiconductor.

Table 1, In-plane elastic modulus (C), effective mass of electrons and holes ($m_e^*, m_h^*$) with respect to the free-electron mass ($m_0$), deformation energy of the CBM and VBM ($E_1^{CBM}$ and $E_1^{VBM}$), and mobility of electrons and holes ($\mu_e, \mu_h$) along zigzag and armchair directions for considered graphyne (GYN) and graphdiyne (GDY) monolayers.

| Structure | Direction | C(N/m) | Electron | | | Hole | | |
|---|---|---|---|---|---|---|---|---|
| | | | $m_e^*$ | $E_1^{CBM}$ | $\mu_e$[1] | $m_h^*$ | $E_1^{VBM}$ | $\mu_h$[1] |
| Pyr-GDY | zigzag | 62.0 | 0.300 | 1.21 | 2142.2 | 0.280 | 6.17 | 1385.6 |
| | armchair | 45.8 | 0.077 | 6.64 | 4146.3 | 0.077 | 1.59 | 8081.9 |
| N-Pyr-GDY | zigzag | 66.3 | 0.380 | 1.04 | 1685.4 | 0.224 | 6.21 | 1874.0 |
| | armchair | 43.3 | 0.082 | 6.38 | 3691.8 | 0.083 | 1.70 | 7749.0 |
| Pyr-GYN | zigzag | 77.4 | 0.423 | 1.86 | 1171.5 | 0.400 | 7.00 | 810.2 |
| | armchair | 61.3 | 0.071 | 7.54 | 3766.8 | 0.072 | 2.03 | 7113.6 |
| N-Pyr-GYN | zigzag | 83.5 | 0.625 | 1.60 | 748.2 | 0.283 | 7.08 | 1350.0 |
| | armchair | 56.9 | 0.075 | 7.21 | 3163.5 | 0.074 | 2.07 | 7846.2 |
| PQ-GYN | zigzag | 228 | 0.045 | 1.35 | 115708.9 | 0.048 | 5.95 | 59004.7 |
| | armchair | 100 | 0.204 | 4.82 | 12409.4 | 0.14 | 0.33 | 35409.5 |

[1]Electron and hole mobilities at 298 K in unit of cm$^2$V$^{-1}$s$^{-1}$.

Finally, we investigate the optical properties of Pyr-, N-Pyr-GYN/GDY and PQ-GYN monolayers by computing their dielectric functions and then absorption coefficients (α(ω)) using the HSE06 functional. For each monolayer, Fig. 6 illustrates two absorption spectra in response to the light incident along the out-of-plane direction and polarized along two planar directions. It can be seen that the monolayers have large absorption coefficients (10$^5$ cm$^{-1}$) in a broad energy range, which are comparable to those of halide perovskites [61]. A moderate anisotropy in both peaks positions and magnitude of absorption coefficients is observed in the absorption spectra of PQ-GYN, however the anisotropy is much weaker for Pyr-, N-Pyr-GYN/GDY. Similar to the case of γ-GDY [51], the first absorption peaks of these monolayers arises from transitions around their band gaps. These peaks are in energy range of 0.26-0.95 eV, which fall in the infrared region and according to our band structure analysis they are due to π→π* transitions. In all cases, strong absorption is also observed in visible and near-UV regions. A broad absorption range and large absorption coefficients make these 2D networks, specifically Pyr- and N-Pyr-GYN/GDN with suitably large band gaps, potential candidates for application in photovoltaic solar cells.

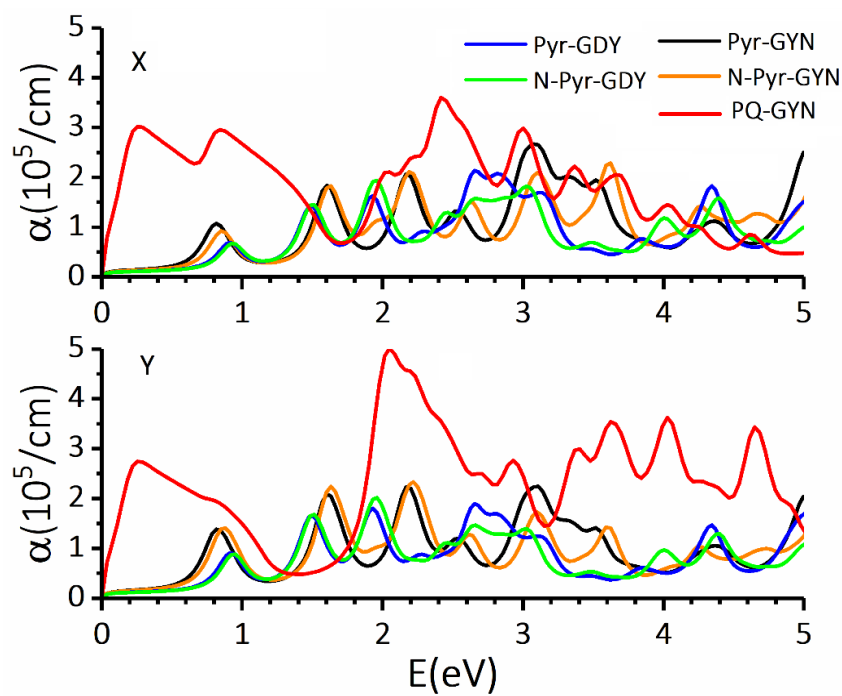

Fig. 6, Absorption coefficients of Pyr-, N-Pyr-GYN/GDY and PQ-GYN monolayers calculated using HSE06 functional. X direction in Pyr- and N-Pyr-GYN/GDY is along the lattice vector of their primitive cells (see Fig. 1), while in PQ-GYN it is along the short diagonal of the rhombic cell (see Fig. 1).

## 4. Concluding remarks

Graphdiyne carbon-based nanoporous nanomaterials are among the most promising candidates to design advanced electronic, optical and energy storage systems. Stimulated by recent synthesis of pyrenyl (Pyr) and pyrazinoquinoxaline(PQ)-based graphdiynes (GDY), using the density functional simulations we investigate thermal and dynamical stability, mechanical, electronic and optical properties of Pyr-, N-Pyr- and PQ-based graphdiyne and graphyne monolayers. All considered nanosheets show desirable thermal stabilities, and their structures could stay completely intact at the elevated temperature of 1000 K. Pyr-based nanosheets are found to show higher stretchability than graphene, whereas PQ counterparts fail at low strain levels.  While Pyr- and N-Pyr-based GYN/GDY and also PQ-GYN are confirmed to be narrow-gap semiconductors, notably we find that PQ-GDY exhibits semimetallic character with distorted Dirac cone and highly anisotropic fermi velocities. The presence of acetylenic linkages in between core molecules facilitates extending the p-conjugation system in these nanosheets, leading to small band gaps, semimetalic states, and ultrasmall carrier effective masses. Fundamental band gaps of semiconducting monolayers are in the range of 0.17 to 0.94 eV, and 0.07 to 0.52 eV within the HSE06 and PBE/GGA functional, respectively. Considered novel 2D systems can exhibit ultrahigh carries mobilities, and specifically in PQ-GYN monolayer, electron and hole mobilities are estimated to be as high as 115708.9 and 59004.7 $cm^2V^{-1}s^{-1}$, distinctly larger than the theoretically estimated value for monolayer phosphorene (10000 $cm^2V^{-1}s^{-1}$). In all cases, electron and hole mobilities are extremely directional, and such that electrons and holes predominantly move along armchair and zigzag directions, respectively. With narrow direct band gaps or Dirac cone and remarkable thermal stability, high mechanical strength, and extremely high electron and hole mobilities, Pyr-, N-Pyr-, and PQ-GYN/GDY could serve as promising candidates for application in nanoelectronics and optoelectronics. Likely to other members of carbon-based graphdiyne family, these novel 2D systems should be naturally attractive for rechargeable metal-ion batteries, thermoelectricity and catalysis as well, which require additional investigations.

## Acknowledgment

F. S. thanks the Persian Gulf University Research Council for support of this study. B.M. appreciates the funding by the Deutsche Forschungsgemeinschaft (DFG, German Research Foundation) under Germany's Excellence Strategy within the Cluster of Excellence PhoenixD (EXC 2122, Project ID 390833453). B. M is also greatly thankful to the VEGAS cluster at Bauhaus University of Weimar for providing the computational resources.

## Appendix A. Supplementary data

The following are the supplementary data to this article:

# Supplementary Materials

# Ultrahigh carrier mobility, Dirac cone and high stretchability in pyrenyl and pyrazinoquinoxaline graphdiyne/graphyne nanosheets confirmed by first-principles

Fazel Shojaei[a] and Bohayra Mortazavi*[,b]

[a]Department of Chemistry, Faculty of Sciences, Persian Gulf University, Bushehr 75169, Iran.
[b]Chair of Computational Science and Simulation Technology, Institute of Photonics, Department of Mathematics and Physics, Leibniz Universität Hannover, Appelstraße 11,30157 Hannover, Germany.

Corresponding author: *bohayra.mortazavi@gmail.com

1. Geometric coordinates of rhombic (rh) and rectangular (Rec.) structures in VASP POSCAR format.

```
Pyr-GYN-rh
  1.00000000000000
    9.7898226308007263   0.5124314261873093   0.0000000000000000
    2.2012428837521174   9.5555975449155550   0.0000000000000000
    0.0000000000000000   0.0000000000000000  15.0000000000000000
   C   H
   20   6
Direct
  0.1671069394302620  0.5716405718209572  0.5000000000000000
  0.3766419601945472  0.3694749731992957  0.5000000000000000
  0.5787542420703033  0.1602592382989698  0.5000000000000000
  0.3504684254485610  0.7597584814313052  0.5000000000000000
  0.4380302026311753  0.8475727257193952  0.5000000000000000
  0.4064293303677999  0.6096872103341369  0.5000000000000000
  0.5568593971635352  0.5498101394382786  0.5000000000000000
  0.6481794720228322  0.6414159298944995  0.5000000000000000
  0.3160883069192479  0.5169173848764714  0.5000000000000000
  0.6169300566818166  0.3994310922619349  0.5000000000000000
  0.7670309536051647  0.3436022144075892  0.5000000000000000
  0.8545846986531203  0.4314005484008376  0.5000000000000000
  0.5241266068495989  0.3091501560625431  0.5000000000000000
  0.6809632289008931  0.8820890429434698  0.5000000000000000
  0.8284506466014534  0.8216714970796914  0.5000000000000000
  0.5881568404179305  0.7917630260781792  0.5000000000000000
  0.8889031285281419  0.6741914914362042  0.5000000000000000
  0.7985750515049972  0.5814199812848386  0.5000000000000000
  0.0378230624652574  0.6194406712499756  0.5000000000000000
  0.6263150682194478  0.0310101069494948  0.5000000000000000
  0.3071725005200818  0.2997767327898728  0.5000000000000000
  0.8980823340418524  0.8912150282684337  0.5000000000000000
  0.2354662538770000  0.8038335067251552  0.5000000000000000
  0.3936781989897753  0.9625118548431004  0.5000000000000000
  0.8113403274873363  0.2286618645436462  0.5000000000000000
  0.9696024654078741  0.3874145436617056  0.5000000000000000
```

```
Pyr-GYN-Rec
   1.00000000000000
     11.7484447178511857    0.0000000000000000    0.0000000000000000
      0.0000000000000000   15.7300172986197673    0.0000000000000000
      0.0000000000000000    0.0000000000000000   15.0000000000000000
   C    H
   40   12
Direct
  0.2854101205231174  0.5502923717360488  0.5000000000000000
  0.3911029234937260  0.6878470341913783  0.5000000000000000
  0.3892530640029079  0.5968099023356430  0.5000000000000000
  0.4947836442421547  0.5515724032734255  0.5000000000000000
  0.7041733327464996  0.5490791043233543  0.5000000000000000
  0.5999219290582047  0.6872765517002719  0.5000000000000000
  0.6008175869439754  0.5962018029945071  0.5000000000000000
  0.7894249460981797  0.7766884392584430  0.5000000000000000
  0.2019918553103039  0.7777230995931177  0.5000000000000000
  0.7853293364846436  0.9618529588207281  0.5000000000000000
  0.7848589643767738  0.0491046250899032  0.5000000000000000
  0.8891984656585095  0.9153336502064491  0.5000000000000000
  0.9947147930424904  0.9605849187216577  0.5000000000000000
  0.9942183545221539  0.0515620280576030  0.5000000000000000
  0.8910715359161614  0.8242716816317284  0.5000000000000000
  0.1007061833226999  0.9158927792753957  0.5000000000000000
  0.2040540346738666  0.9630068226177784  0.5000000000000000
  0.2036013039469040  0.0502703706698142  0.5000000000000000
  0.0998794880701581  0.8248312185720010  0.5000000000000000
  0.8891244799980669  0.1872835478943387  0.5000000000000000
  0.8882180892730673  0.0962105659834762  0.5000000000000000
  0.0979201925779734  0.1878283082948471  0.5000000000000000
  0.0997561079447706  0.0967879951670199  0.5000000000000000
  0.7018993978456578  0.7345191661500508  0.5000000000000000
  0.2894996056261121  0.7355248324772461  0.5000000000000000
  0.2870054974327729  0.2777294350797064  0.5000000000000000
  0.2849591179889970  0.4630223880220043  0.5000000000000000
  0.3883000451383865  0.4158999264240109  0.5000000000000000
  0.3891326365027012  0.3248333777393952  0.5000000000000000
  0.1995139342052283  0.2355128326285154  0.5000000000000000
  0.6996001538830754  0.2766647398855824  0.5000000000000000
  0.4942886317596873  0.4605863613925649  0.5000000000000000
  0.5998063106695497  0.4153205054560374  0.5000000000000000
  0.7036859194109866  0.4618266391854249  0.5000000000000000
  0.5979470705654251  0.3242544970117720  0.5000000000000000
  0.7871274497210266  0.2344952758390093  0.5000000000000000
  0.4957041156445428  0.7308297038612395  0.5000000000000000
  0.9956608632025166  0.7812383209673541  0.5000000000000000
  0.9933357951422295  0.2308283932787845  0.5000000000000000
  0.4933473693049280  0.2812286844459848  0.5000000000000000
  0.2056049204115808  0.5857638995787653  0.5000000000000000
  0.7843294602822297  0.5841164586900334  0.5000000000000000
  0.7055243126440374  0.9263878747607919  0.5000000000000000
  0.7047133871712603  0.0841557966728743  0.5000000000000000
  0.2842135184541945  0.9279768368548176  0.5000000000000000
  0.2834162493634764  0.0857325559188098  0.5000000000000000
  0.2047889016501330  0.4280093861156544  0.5000000000000000
  0.7834828613149512  0.4263491337245284  0.5000000000000000
  0.4960151782029669  0.8001005136761918  0.5000000000000000
  0.9959312680904802  0.7119825753470295  0.5000000000000000
  0.9930483189446662  0.3000989897759609  0.5000000000000000
  0.4930267301989062  0.2119727986309314  0.5000000000000000
```

```
N-Pyr-GYN-rh
 1.00000000000000
  9.6992386886183990   0.0000000000000000   0.0000000000000000
  2.7070338264525948   9.3160582205445035   0.0000000000000000
  0.0000000000000000   0.0000000000000000  15.0000000000000000
  C  H  N
 18  4  2
Direct
 0.1672611244967749 0.5666745827233084 0.5000000000000000
 0.5737463411632781 0.1603352049091148 0.5000000000000000
 0.3473718744086489 0.7618707329495292 0.5000000000000000
 0.4359761950554513 0.8505993278523204 0.5000000000000000
 0.4052838578472248 0.6098361826409281 0.5000000000000000
 0.5567174920321944 0.5497191725225092 0.5000000000000000
 0.6483894078430188 0.6415011385462965 0.5000000000000000
 0.3189017861094430 0.5115474055359529 0.5000000000000000
 0.6169153655872670 0.3982915418576822 0.5000000000000000
 0.7690407985836387 0.3405610903480404 0.5000000000000000
 0.8576569140644208 0.4292540738747114 0.5000000000000000
 0.5186688744009301 0.3119158853148884 0.5000000000000000
 0.6863756375839729 0.8792998416794778 0.5000000000000000
 0.5880982436138541 0.7929211446680540 0.5000000000000000
 0.8861731311239610 0.6796164439411362 0.5000000000000000
 0.7998297371427354 0.5812973584463053 0.5000000000000000
 0.0378302222510618 0.6245220258462085 0.5000000000000000
 0.6313014305050716 0.0309015087736251 0.5000000000000000
 0.2309533060651654 0.8068276234126877 0.5000000000000000
 0.3908572331847435 0.9669668524953144 0.5000000000000000
 0.8140652379472352 0.2241933012924852 0.5000000000000000
 0.9740565934052272 0.3843026237185460 0.5000000000000000
 0.3746846673267378 0.3676320483741600 0.5000000000000000
 0.8303494672579326 0.8235405542767336 0.5000000000000000
```

N-Pyr-GYN-Rec
   1.00000000000000
     11.5500954528944764    0.0000000000000000    0.0000000000000000
      0.0000000000000000   15.6326423901666196    0.0000000000000000
      0.0000000000000000    0.0000000000000000   15.0000000000000000
    C    H    N
   36    8    4
Direct
  0.2821399229485593  0.5507814369146473  0.5000000000000000
  0.3949011407207479  0.6887291313097990  0.5000000000000000
  0.3882864142931410  0.5972116911390373  0.5000000000000000
  0.4947673101802010  0.5516740655897507  0.5000000000000000
  0.7073536926906385  0.5495561972606779  0.5000000000000000
  0.5960099489942081  0.6881667213252172  0.5000000000000000
  0.6017308500098153  0.5966176898569344  0.5000000000000000
  0.7918272102450459  0.7747254281488907  0.5000000000000000
  0.1995746806888619  0.7758601766165114  0.5000000000000000
  0.7820236873577500  0.9616157647169601  0.5000000000000000
  0.7816278400515273  0.0496091010591471  0.5000000000000000
  0.8881758963577084  0.9151483324601770  0.5000000000000000
  0.9946755118721811  0.9606933654265148  0.5000000000000000
  0.9942492202545665  0.0516892074811324  0.5000000000000000
  0.8948998642352706  0.8236196739300894  0.5000000000000000
  0.1016512933656486  0.9157257328095838  0.5000000000000000
  0.2072884130644965  0.9627782400726232  0.5000000000000000
  0.2068681096122802  0.0507683097521578  0.5000000000000000
  0.0960314306679209  0.8241968900238987  0.5000000000000000
  0.8930456520822290  0.1881511477371944  0.5000000000000000
  0.8873063873047187  0.0966294935308341  0.5000000000000000
  0.0941591627482410  0.1887132898905293  0.5000000000000000
  0.1007477232649805  0.0972081186989442  0.5000000000000000
  0.6995043785636170  0.7365747567775358  0.5000000000000000
  0.2918255415570812  0.7376136847438417  0.5000000000000000
  0.2895248399935113  0.2757905782811605  0.5000000000000000
  0.2816560291585475  0.4627835791416075  0.5000000000000000
  0.3873130801892941  0.4156985036910115  0.5000000000000000
  0.3930241525331510  0.3241552947913533  0.5000000000000000
  0.1972800092049596  0.2375586947060526  0.5000000000000000
  0.6972884600405678  0.2747650076806707  0.5000000000000000
  0.4942682641776415  0.4606627405652048  0.5000000000000000
  0.6007714850658985  0.4151112990917554  0.5000000000000000
  0.7069058009572515  0.4615662701287633  0.5000000000000000
  0.5941297523704705  0.3236018730929260  0.5000000000000000
  0.7895781877886492  0.2365877186003473  0.5000000000000000
  0.2010631037391093  0.5866105819600946  0.5000000000000000
  0.7888444658656866  0.5848635116246115  0.5000000000000000
  0.7008891521979166  0.9258686174761834  0.5000000000000000
  0.7001638400753123  0.0849551288552490  0.5000000000000000
  0.2887813809742568  0.9274867900541395  0.5000000000000000
  0.2879695491837069  0.0865610674745554  0.5000000000000000
  0.2001315559020753  0.4275315677334888  0.5000000000000000
  0.7880283759028188  0.4258065251647380  0.5000000000000000
  0.4956747234611427  0.7322245135664005  0.5000000000000000
  0.9956920273982774  0.7801426347369045  0.5000000000000000
  0.9933850562531106  0.2322118173568484  0.5000000000000000
  0.4933654084351531  0.2801079279533170  0.5000000000000000

```
PQ-GYN-rh
1.0
    11.1431999207     0.0000000000     0.0000000000
     8.9893594591     6.5828074193     0.0000000000
     0.0000000000     0.0000000000    14.0000000000
  C   N
 16   4
Direct
   0.010740058    0.431640002    0.500000000
   0.977639980    0.566170000    0.500000000
   0.582730013    0.410530014    0.500000000
   0.405080031    0.588039994    0.500000000
   0.427150029    0.015000000    0.500000000
   0.561049997    0.982770069    0.500000000
   0.932180158    0.727030032    0.500000000
   0.721859948    0.937380044    0.500000000
   0.039549954    0.837040010    0.500000000
   0.831800000    0.044860001    0.500000000
   0.204390095    0.788630024    0.500000000
   0.155950013    0.953509969    0.500000000
   0.948200011    0.161359990    0.500000000
   0.266250028    0.060540003    0.500000000
   0.055989988    0.270930005    0.500000000
   0.783329971    0.209830007    0.500000000
   0.084859939    0.681930043    0.500000000
   0.311180036    0.907970000    0.500000000
   0.676760059    0.090080003    0.500000000
   0.903119941    0.316309979    0.500000000
```

PQ-GYN-Rec
1.0
    21.1833992004     0.0000000000     0.0000000000
     0.0000000000     6.9099998474     0.0000000000
     0.0000000000     0.0000000000    14.0000000000
  C  N
 31  8
Direct
   0.999790027    0.210879996    0.500000000
   0.999759954    0.410919985    0.500000000
   0.058079998    0.103990001    0.500000000
   0.058149998    0.896340016    0.500000000
   0.166649994    0.105660000    0.500000000
   0.166710006    0.894989966    0.500000000
   0.224480002    0.791769970    0.500000000
   0.224440002    0.208640000    0.500000000
   0.275239997    0.708289997    0.500000000
   0.275319997    0.291339988    0.500000000
   0.333039999    0.605169991    0.500000000
   0.333079999    0.394490012    0.500000000
   0.441610012    0.603519969    0.500000000
   0.441620007    0.395889988    0.500000000
   0.499879977    0.710420031    0.500000000
   0.499839999    0.288670011    0.500000000
   0.499820010    0.088610001    0.500000000
   0.499910005    0.910449980    0.500000000
   0.558099970    0.395690006    0.500000000
   0.558120004    0.603319987    0.500000000
   0.666689972    0.604889962    0.500000000
   0.666649994    0.394190005    0.500000000
   0.724510008    0.707850010    0.500000000
   0.724399979    0.290980014    0.500000000
   0.775190024    0.207770013    0.500000000
   0.775360020    0.790769992    0.500000000
   0.832999976    0.104860004    0.500000000
   0.833100010    0.894190039    0.500000000
   0.941559950    0.103770007    0.500000000
   0.941649990    0.896130028    0.500000000
  -0.000090040    0.789149988    0.500000000
   0.113020004    0.795729996    0.500000000
   0.112949999    0.204810000    0.500000000
   0.386750034    0.295230007    0.500000000
   0.386739994    0.704320034    0.500000000
   0.612990021    0.704079960    0.500000000
   0.612959993    0.294979995    0.500000000
   0.886650000    0.204289999    0.500000000
   0.886819995    0.795210029    0.500000000

Pyr-GDY-rh
1.0
    12.3281698227     0.0000000000     0.0000000000
     2.1236638294    12.1330986433     0.0000000000
     0.0000000000     0.0000000000    15.0000000000
  C   H
 24   6
Direct
    0.165329006    0.462065991    0.500000000
    0.321275003    0.313405994    0.500000000
    0.469704988    0.157048999    0.500000000
    0.317604017    0.619723981    0.500000000
    0.390475010    0.692458997    0.500000000
    0.355264970    0.503955007    0.500000000
    0.471050006    0.462795998    0.500000000
    0.546988997    0.538574997    0.500000000
    0.280160025    0.427075003    0.500000000
    0.511868026    0.346890966    0.500000000
    0.627499015    0.308878986    0.500000000
    0.700372974    0.381594981    0.500000000
    0.434796962    0.271961978    0.500000000
    0.548593998    0.844278972    0.500000000
    0.583286007    0.729363055    0.500000000
    0.696773026    0.687898014    0.500000000
    0.506138050    0.654511981    0.500000000
    0.737880993    0.574225015    0.500000000
    0.852648082    0.539220980    0.500000000
    0.662744029    0.497377985    0.500000000
    0.953736965    0.514318988    0.500000000
    0.523184013    0.945416990    0.500000000
    0.064231001    0.487100976    0.500000000
    0.495332978    0.055929998    0.500000000
    0.263496006    0.255721979    0.500000000
    0.754619011    0.745518951    0.500000000
    0.229009002    0.649810027    0.500000000
    0.360623027    0.781170031    0.500000000
    0.657369990    0.220163001    0.500000000
    0.788968939    0.351438980    0.500000000

```
Pyr-GDN-Rec
1.0
    15.8605003357     0.0000000000     0.0000000000
     0.0000000000    18.8589992523     0.0000000000
     0.0000000000     0.0000000000    15.0000000000
  C  H
 48  12
Direct
   0.499709998    0.536559978    0.500000000
   0.654799969    0.534460013    0.500000000
   0.500429982    0.686230012    0.500000000
   0.577610045    0.649749983    0.500000000
   0.652559988    0.689580034    0.500000000
   0.578280001    0.573730016    0.500000000
   0.715650047    0.727580005    0.500000000
   0.344619997    0.535629970    0.500000000
   0.348140003    0.690689966    0.500000000
   0.422829978    0.650390031    0.500000000
   0.421519980    0.574349988    0.500000000
   0.284870007    0.728529988    0.500000000
   0.847640001    0.807149994    0.500000000
   0.999849978    0.811079988    0.500000000
   0.152149999    0.807370018    0.500000000
   0.844479987    0.962260002    0.500000000
   0.844409997    0.035099982    0.500000000
   0.921249971    0.923259935    0.500000000
   0.999589981    0.960760035    0.500000000
   0.999499968    0.036669971    0.500000000
   0.922410038    0.847230006    0.500000000
   0.078009962    0.923349947    0.500000000
   0.154660004    0.962430014    0.500000000
   0.154580003    0.035259962    0.500000000
   0.077190045    0.847299993    0.500000000
   0.847159992    0.190249993    0.500000000
   0.922000019    0.150199994    0.500000000
   0.999380011    0.186369999    0.500000000
   0.921079987    0.074139951    0.500000000
   0.076750036    0.150199994    0.500000000
   0.151490009    0.190290006    0.500000000
   0.077819940    0.074190020    0.500000000
   0.214520000    0.228349989    0.500000000
   0.783909989    0.228100003    0.500000000
   0.784560043    0.769140010    0.500000000
   0.215690003    0.769849994    0.500000000
   0.498760022    0.310970001    0.500000000
   0.651130002    0.306840008    0.500000000
   0.499359988    0.460660010    0.500000000
   0.577530013    0.422899987    0.500000000
   0.654410033    0.461629967    0.500000000
   0.576300017    0.346870007    0.500000000
   0.714630021    0.269279984    0.500000000
   0.346599982    0.307749991    0.500000000
   0.344269987    0.462799975    0.500000000
   0.420819990    0.423500010    0.500000000
   0.421559996    0.347499992    0.500000000
   0.283430011    0.269870019    0.500000000
   0.500840001    0.743989991    0.500000000
   0.714269969    0.563549989    0.500000000
   0.285520001    0.565230031    0.500000000
   0.999809993    0.753319959    0.500000000
   0.999440020    0.244129990    0.500000000
   0.785210036    0.932890028    0.500000000
   0.785109982    0.064450029    0.500000000
   0.213979998    0.933120014    0.500000000
   0.213850014    0.064630034    0.500000000
   0.498350003    0.253219985    0.500000000
   0.713549986    0.432040020    0.500000000
   0.284829992    0.433659987    0.500000000
```

N-Pyr-GDN-rh
1.0
    12.2153501511     0.0000000000     0.0000000000
     2.0836270596    12.0285673332     0.0000000000
     0.0000000000     0.0000000000    15.0000000000
  C   H   N
 22   4   2
Direct
    0.165915988     0.455770999     0.500000000
    0.463607038     0.157639992     0.500000000
    0.316213972     0.620190040     0.500000000
    0.389915980     0.693801971     0.500000000
    0.354662026     0.503368979     0.500000000
    0.470908983     0.462630978     0.500000000
    0.547128957     0.538752965     0.500000000
    0.282446006     0.422209993     0.500000000
    0.511348009     0.346296999     0.500000000
    0.628064954     0.307577016     0.500000000
    0.701799987     0.381199016     0.500000000
    0.430063972     0.274203993     0.500000000
    0.554535964     0.843747974     0.500000000
    0.587947013     0.727156957     0.500000000
    0.506639998     0.655116040     0.500000000
    0.735599018     0.579155991     0.500000000
    0.852045985     0.545435030     0.500000000
    0.663361985     0.498033000     0.500000000
    0.953435001     0.515956998     0.500000000
    0.525161021     0.945205002     0.500000000
    0.064475987     0.484885989     0.500000000
    0.493609003     0.056250999     0.500000000
    0.226840005     0.650350986     0.500000000
    0.359919034     0.783254036     0.500000000
    0.658070055     0.218121998     0.500000000
    0.791165949     0.350984990     0.500000000
    0.319724997     0.311738005     0.500000000
    0.698301978     0.689620051     0.500000000

N-Pyr-GDY-Rec
1.0
    15.7060804367     0.0000000000     0.0000000000
     0.0000000000    18.6896800995     0.0000000000
     0.0000000000     0.0000000000    15.0000000000
  C   H   N
 44   8   4
Direct
   0.411450008    0.440247990    0.500000000
   0.567846027    0.438559004    0.500000000
   0.485864003    0.554422007    0.500000000
   0.561016051    0.595781003    0.500000000
   0.490166020    0.477821003    0.500000000
   0.626347022    0.631936033    0.500000000
   0.255094004    0.439494018    0.500000000
   0.262943014    0.596591002    0.500000000
   0.337896993    0.554898036    0.500000000
   0.333050013    0.478307033    0.500000000
   0.197502004    0.632646020    0.500000000
   0.762606990    0.708243020    0.500000000
   0.060812999    0.708545048    0.500000000
   0.754960025    0.865361946    0.500000000
   0.754888011    0.938994007    0.500000000
   0.832834002    0.826345028    0.500000000
   0.911331027    0.864217007    0.500000000
   0.911214991    0.940314988    0.500000000
   0.837612975    0.749758005    0.500000000
   0.989942024    0.826491015    0.500000000
   0.067684003    0.865650043    0.500000000
   0.067575997    0.939270980    0.500000000
   0.985580015    0.749883021    0.500000000
   0.761998028    0.096173999    0.500000000
   0.837100012    0.054700001    0.500000000
   0.832631015    0.978101038    0.500000000
   0.985021998    0.054788003    0.500000000
   0.059999001    0.096353001    0.500000000
   0.989689003    0.978207990    0.500000000
   0.125244004    0.132609000    0.500000000
   0.696564015    0.132214996    0.500000000
   0.697293992    0.672069009    0.500000000
   0.126381996    0.672658011    0.500000000
   0.559753984    0.207925008    0.500000000
   0.411161011    0.364151004    0.500000000
   0.489552990    0.326103977    0.500000000
   0.567537994    0.364925998    0.500000000
   0.484739012    0.249511010    0.500000000
   0.625341012    0.172076994    0.500000000
   0.261633981    0.208754997    0.500000000
   0.254777986    0.365874000    0.500000000
   0.332455018    0.326588987    0.500000000
   0.336832995    0.249997015    0.500000000
   0.196180993    0.172760001    0.500000000
   0.627781958    0.468061008    0.500000000
   0.195433013    0.469392960    0.500000000
   0.695161989    0.835662009    0.500000000
   0.695046013    0.968643019    0.500000000
   0.127541006    0.836032974    0.500000000
   0.127333002    0.969033986    0.500000000
   0.627206028    0.335039992    0.500000000
   0.194821000    0.336398989    0.500000000
   0.412034985    0.591171958    0.500000000
   0.911651022    0.713307021    0.500000000
   0.911024998    0.091271996    0.500000000
   0.410629012    0.213218004    0.500000000

PQ-GDY-rh
1.0
    13.6819200516    0.0000000000    0.0000000000
    10.3803393436    8.9038486932    0.0000000000
     0.0000000000    0.0000000000   14.0000000000
  C   N
 22   4
Direct
    0.913757015    0.812915990    0.500000000
    0.760167105    0.966563952    0.500000000
    0.002764984    0.903549044    0.500000000
    0.850833975    0.055582999    0.500000000
    0.128483054    0.874223007    0.500000000
    0.099384103    0.999846937    0.500000000
    0.947494974    0.151933001    0.500000000
    0.190497985    0.088386997    0.500000000
    0.036933006    0.242180010    0.500000000
    0.675746972    0.327367018    0.500000000
    0.545325031    0.457700974    0.500000000
    0.821700942    0.181292000    0.500000000
    0.404944978    0.597945964    0.500000000
    0.274468036    0.728196964    0.500000000
    0.637911996    0.994220977    0.500000000
    0.313049995    0.060268998    0.500000000
    0.418094007    0.039025000    0.500000000
    0.533015007    0.015247000    0.500000000
    0.009410009    0.364380980    0.500000000
    0.988305968    0.469375971    0.500000000
    0.963788968    0.584820990    0.500000000
    0.941691066    0.690462995    0.500000000
    0.030534984    0.785485981    0.500000000
    0.217589044    0.971780974    0.500000000
    0.732782005    0.083388998    0.500000000
    0.919987971    0.269845014    0.500000000

```
PQ-GDY-Rec.
1.0
    25.6506900787    0.0000000000    0.0000000000
     0.0000000000    9.4740400314    0.0000000000
     0.0000000000    0.0000000000   14.0000000000
  C  N
 43  8
Direct
 0.0017261930059668 0.9993592153124737 0.5000000000000000
 0.0499967436700200 0.0769738787051068 0.5000000000000000
 0.0500751683617642 0.2287551333371183 0.5000000000000000
 0.0018960617209518 0.3068286260325099 0.5000000000000000
 0.0020524361825113 0.4523240457349517 0.5000000000000000
 0.0020752047042597 0.5831037763545217 0.5000000000000000
 0.0019753527749570 0.7230578975311398 0.5000000000000000
 0.0017506209896085 0.8538469370903372 0.5000000000000000
 0.1870102949253267 0.9999524329795919 0.5000000000000000
 0.2288071513685683 0.9364230054619491 0.5000000000000000
 0.2743939778983158 0.8667427878327488 0.5000000000000000
 0.3163999720298101 0.8041296236366746 0.5000000000000000
 0.3637877955304573 0.7290740442222307 0.5000000000000000
 0.3638199270618787 0.5752102786970477 0.5000000000000000
 0.3165053173377700 0.4999793304694153 0.5000000000000000
 0.2746090586689741 0.4369950843137537 0.5000000000000000
 0.2290418630922772 0.3673359085982842 0.5000000000000000
 0.1871534255527636 0.3042994187450699 0.5000000000000000
 0.1398023899812143 0.2292649862086138 0.5000000000000000
 0.1397276802329507 0.0754160128932438 0.5000000000000000
 0.5021146818748697 0.0819156337481364 0.5000000000000000
 0.5021318741535129 0.2218503173171128 0.5000000000000000
 0.5019406327843529 0.3526592920096864 0.5000000000000000
 0.5018069639448026 0.4982150254628750 0.5000000000000000
 0.4535474247037691 0.5760657392068520 0.5000000000000000
 0.4535401887579409 0.7279323053930699 0.5000000000000000
 0.5018052105467499 0.8056979549344961 0.5000000000000000
 0.5500430962129025 0.7279605787770888 0.5000000000000000
 0.5500193111549545 0.5761597620518788 0.5000000000000000
 0.5019042408389313 0.9511544705615459 0.5000000000000000
 0.6397854026935377 0.7291424093324669 0.5000000000000000
 0.6397264705597291 0.5752878593287000 0.5000000000000000
 0.6871434696818876 0.8043633017403667 0.5000000000000000
 0.7290282446924833 0.8675159930693404 0.5000000000000000
 0.7745546326959892 0.9375583613578501 0.5000000000000000
 0.8163807480687048 0.0010409879987596 0.5000000000000000
 0.6869897001521589 0.4999764263890410 0.5000000000000000
 0.7288211631860346 0.4367265911913656 0.5000000000000000
 0.7744340447115469 0.3673949950963314 0.5000000000000000
 0.8165075554697563 0.3052786669571249 0.5000000000000000
 0.8638320878505621 0.2301713631192748 0.5000000000000000
 0.8637391380422470 0.0763004333796540 0.5000000000000000
 0.9535007983823958 0.0773286075437863 0.5000000000000000
 0.9535765359027198 0.2291725935731186 0.5000000000000000
 0.0952576950767991 0.3018076962786946 0.5000000000000000
 0.4084132889724462 0.5028321156624269 0.5000000000000000
 0.4084207300872622 0.8013067261339309 0.5000000000000000
 0.5951633211322636 0.8013548154304073 0.5000000000000000
 0.5951432047760578 0.5029040454011915 0.5000000000000000
 0.9084559167097891 0.3024672641377128 0.5000000000000000
 0.9083570469380541 0.0040334530195148 0.5000000000000000
 0.0950641601545242 0.0033892762394287 0.5000000000000000
```

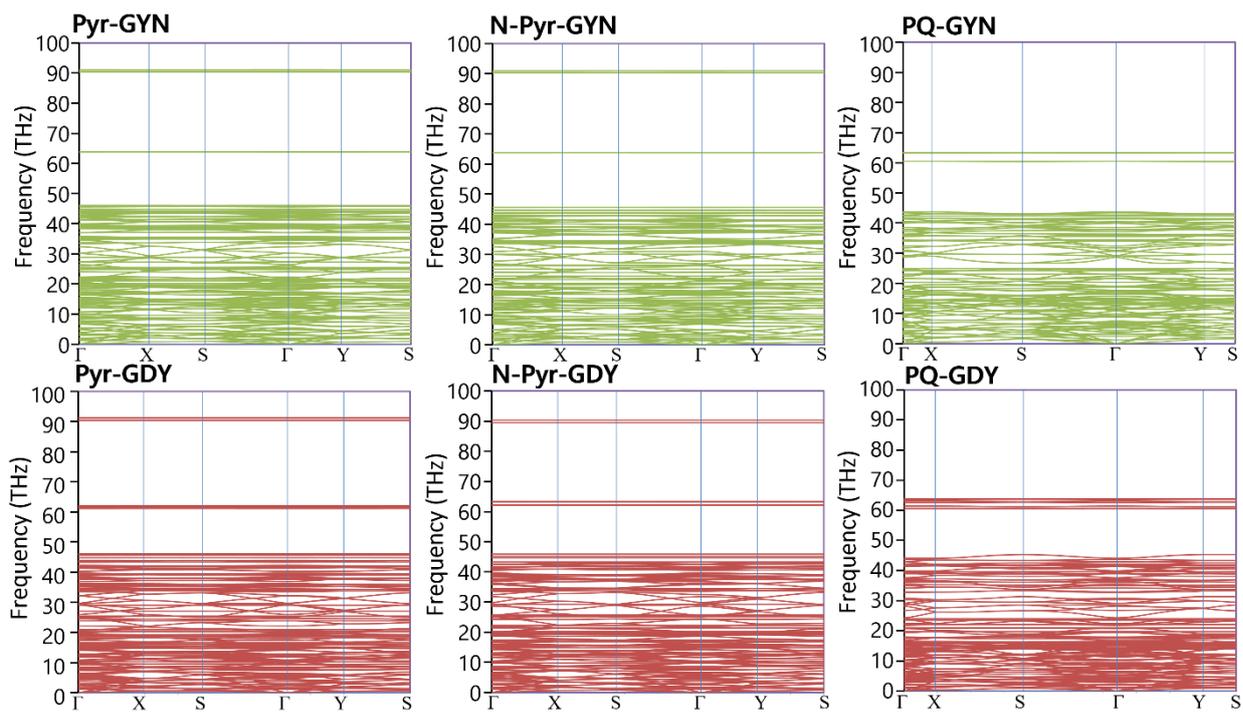

**Fig. S1**, Complete phonon dispersion relations up to 100 THz.

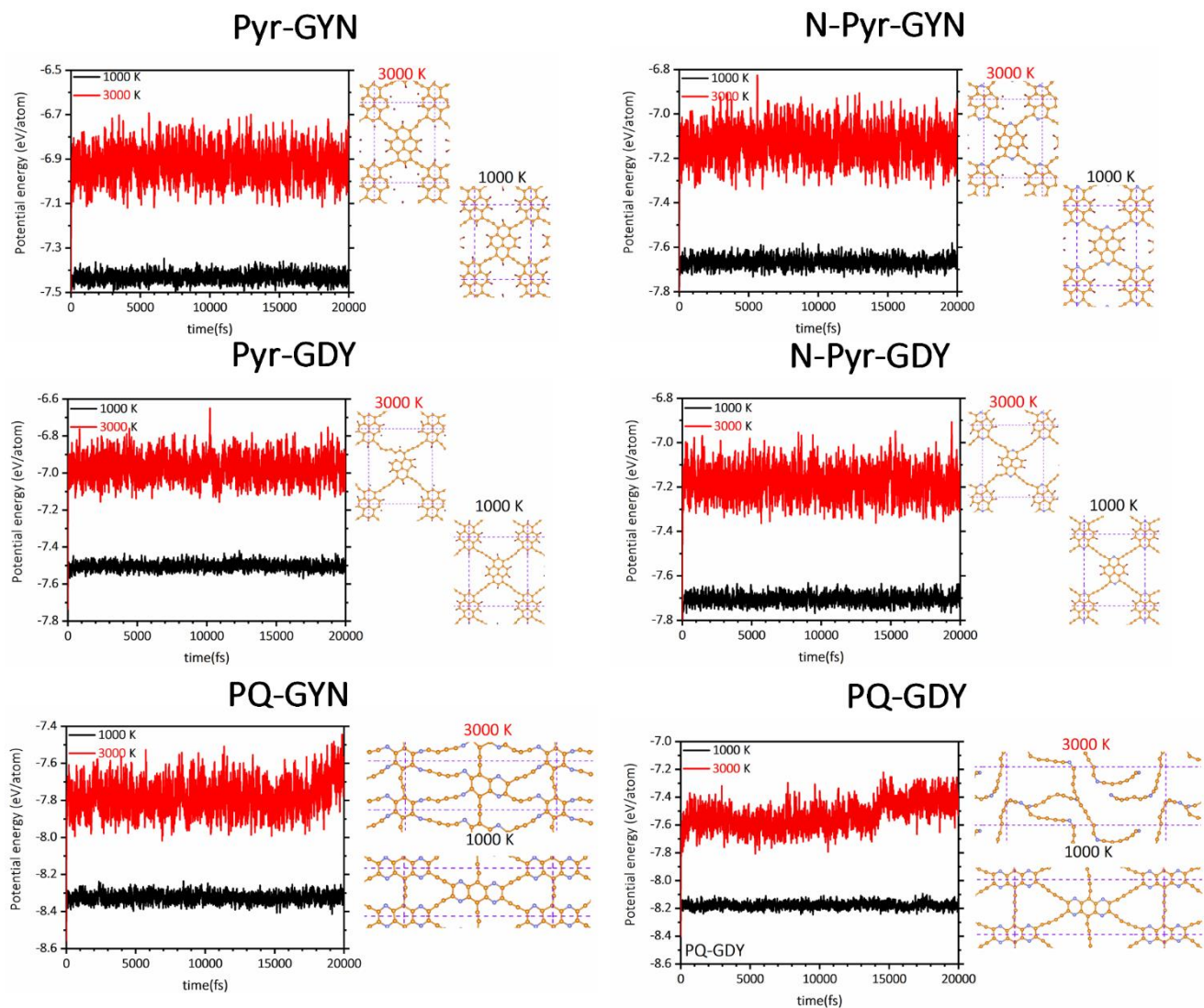

Fig. S2, Fluctuation of potential energy and top view of rectangular supercell of Pyr-, N-Pyr, and PQ-GYN/GDY after 20 ps AIMD simulations at 1000, and 3000 K.

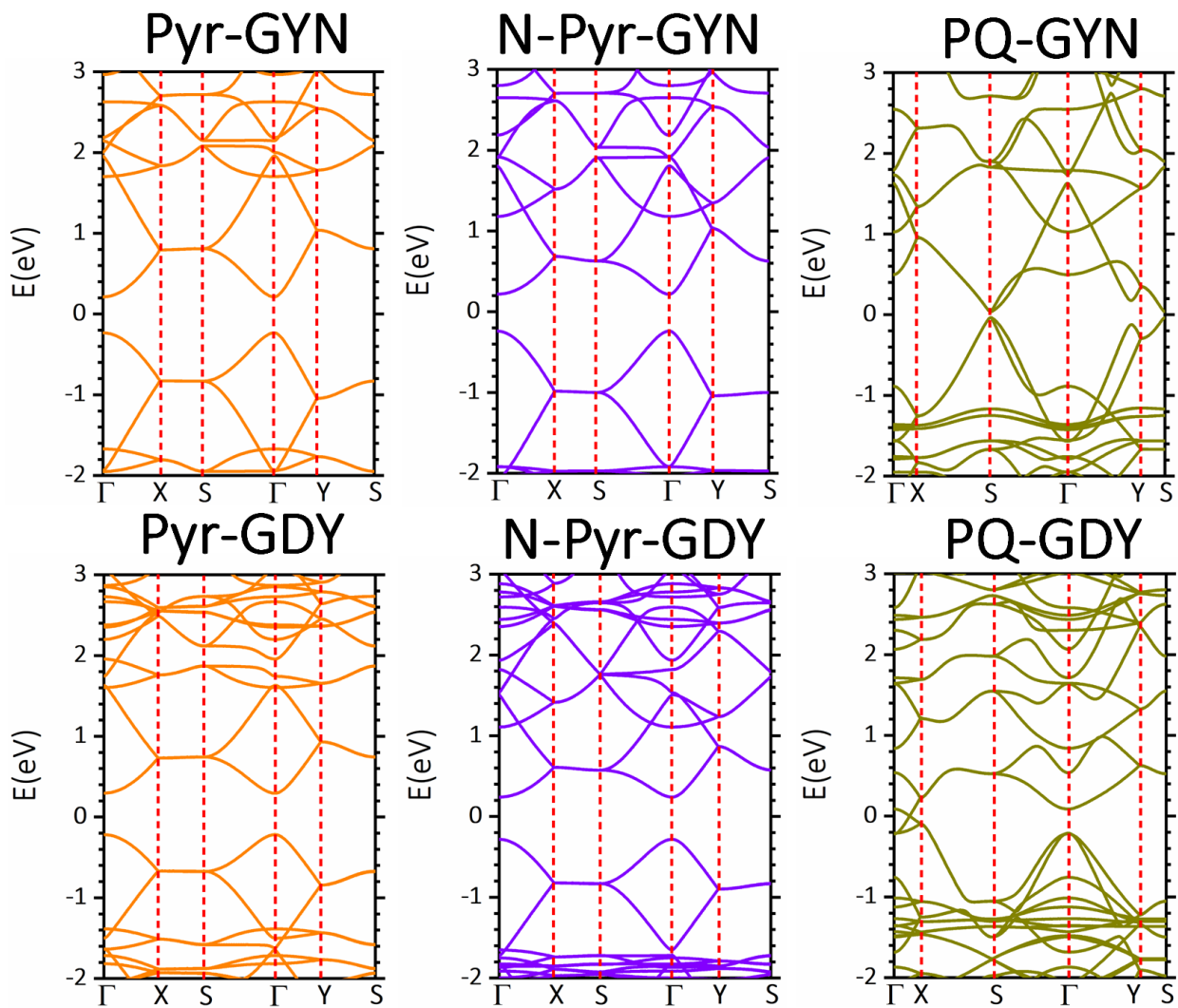

Fig. S3, PBE/GGA band structure of rectangular Pyr-,N-Pyr, and PQ-GYN/GDY monolayers.